\documentstyle[12pt,epsf]{article}

\newcommand{\bmat}{\left(\begin{array}}
\newcommand{\emat}{\end{array}\right)}
\def\NPB#1#2#3{Nucl. Phys. B{#1} (19#2) #3}
\def\PLB#1#2#3{Phys. Lett. B{#1} (19#2) #3}

\def\PRD#1#2#3{Phys. Rev. D{#1} (19#2) #3}

\def\Deq#1{\mbox{$D$=#1}}
\def\Neq#1{\mbox{$N$=#1}}

\def\yzero{\smash{\hbox{$y\kern-4pt\raise1pt\hbox{${}^\circ$}$}}}

\def\g{\gamma}

\def\-{\hphantom{-}}
\def\ov{\overline}
\def\s2{\frac{1}{\sqrt2}}

\def\beq{\begin{equation}}
\def\eeq{\end{equation}}
\def\beqa{\begin{eqnarray}}
\def\eeqa{\end{eqnarray}}

\def\Tr{{\rm Tr \,}}
\def\diag{{\rm diag \,}}
\def\tg{{\rm tg \,}}
\def\cotg{{\rm cotg \,}}
\def\IF{\relax{\rm I\kern-.18em F}}
\def\II{\relax{\rm I\kern-.18em I}}
\def\IP{\relax{\rm I\kern-.18em P}}
\def\inbar{\vrule height1.5ex width.4pt depth0pt}
\def\IC{\relax\hbox{\kern.25em$\inbar\kern-.3em{\rm C}$}}
\def\IR{\relax{\rm I\kern-.18em R}}

\def\Dsl{\,\raise.15ex\hbox{/}\mkern-13.5mu D} 

\topmargin
-1.5cm
\textwidth
15.5cm
\textheight
24cm
\oddsidemargin
0.7cm
\evensidemargin
1.2cm

\begin{document}

\makeatletter
\@addtoreset{equation}{section}
\makeatother
\renewcommand{\theequation}{\thesection.\arabic{equation}}
\pagestyle{empty}
\rightline{FTUAM-99/14, IASSNS-HEP-99/48, IFT-UAM/CSIC-99-17}
\rightline{\tt hep-th/yymmddd}
\vspace{0.5cm}
\begin{center}
\LARGE{ Sigma-model anomalies in compact  \\
 D=4, N=1 Type IIB orientifolds
and Fayet-Iliopoulos terms\\[10mm]}
\large{
L.~E.~Ib\'a\~nez$^1$, R. Rabad\'an$^1$
and A.~M.~Uranga$^2$\\[2mm]}
\small{
$^1$ Departamento de F\'{\i}sica Te\'orica C-XI
and Instituto de F\'{\i}sica Te\'orica  C-XVI,\\[-0.3em]
Universidad Aut\'onoma de Madrid,
Cantoblanco, 28049 Madrid, Spain.\\[4mm]
$^2$ Institute for Advanced Study, Olden Lane, Princeton NJ 08540,
USA.\\[4mm]
 }
\small{\bf Abstract} \\[7mm]
\end{center}

\begin{center}
\begin{minipage}[h]{14.0cm}

{\small
Compact Type IIB $D=4$, $N=1$ orientifolds have certain $U(1)$
$\sigma$-model
symmetries at the level of the effective Lagrangian. These symmetries are
generically anomalous. We study the particular case of $Z_N$ orientifolds
and find that these anomalies may be cancelled by a generalized Green-Schwarz
mechanism. This mechanism works by the exchange of twisted RR-fields
associated to the orbifold singularities and it requires the mixing between
twisted and untwisted moduli of the orbifold. As a consequence, the
Fayet-Iliopoulos terms which are present for the gauged anomalous $U(1)$'s
of the models get an additional untwisted modulus dependent piece at the
tree level.
}

\end{minipage}
\end{center}
\newpage
\setcounter{page}{1}
\pagestyle{plain}
\renewcommand{\thefootnote}{\arabic{footnote}}
\setcounter{footnote}{0}

\section{Introduction}

The low-energy dynamics of moduli fields in string theory is described by
a non-linear $\sigma$-model. The isometries of the corresponding target
space appear as symmetries of the effective lagrangian. In general it is
expected that these symmetries are not preserved by full-fledged string
theory, since the properties of states in the massive tower depend
non-trivially on the moduli. However, quite often a discrete version of
these symmetries is valid for the complete string theory, including the
massive stringy modes (or even non-perturbative states). The study of
$\sigma$-model symmetries thus may provide interesting insights into
deeper properties of string theory.

A simple example is provided by $D=4$ $N=1$ heterotic orbifold vacua. The
classical low-energy lagrangian is invariant under a number of
$SL(2,{\bf R})$ transformations acting on the untwisted moduli $T_i$
controlling the sizes of the compact dimensions. The masses of momentum
and winding modes depend on these moduli, so the continuous symmetry is
violated in the full string theory. However, the corresponding discrete
$SL(2,{\bf Z})$ modular transformations correspond to an exact symmetry of
the full theory, T-duality.

A new ingredient comes about when one realizes that the $\sigma$-model
symmetries involve a non-trivial transformation of chiral fermions charged
under gauge symmetries. This leads to potential $\sigma$-gauge (and
$\sigma$-gravity) mixed anomalies, spoiling the quantum validity of the
symmetry. In fact, direct computation shows that the triangle diagram
contributions give a non-vanishing anomaly. Even though this is not of
great concern for the continuous version of the symmetry, which is anyway
broken by other effects, such anomalies would be clearly inconsistent
with T-duality
being an exact symmetry of the theory. Happily, the triangle contribution
is cancelled by additional effects. First, the gauge kinetic functions
have a non-trivial one-loop dependence (threshold correction \cite{kaplu,dkl})
on the untwisted moduli associated  to complex planes left unrotated by some
orbifold group element. Second, a Green-Schwarz mechanism \cite{gs} with
dilaton exchange cancels the remaining  anomaly \cite{dfkz,lco}. Notice
that for complex planes rotated by all orbifold group elements, there is
no threshold correction \cite{dkl}, so the GS counterterms cancel the
anomalies not only for the discrete but also for the continuous version
of the $\sigma$-model symmetry.

\medskip

It is natural to consider similar questions for other $D=4$ $N=1$ vacua.
In the present paper we center on type IIB orientifolds
\cite{bl} -\cite{iru} . The low-energy
effective lagrangians for moduli are quite analogous to those of heterotic
models, for instance they have $SL(2,{\bf R})$ $\sigma$-model symmetries
for the untwisted moduli $T_i$. An important difference, however, is that in
these type of vacua T-duality is not related to these modular
transformations. In fact, T-duality relates D-branes of different kinds,
and so does not act within a given class of models, but maps one class of
vacua to another. For instance, an orientifold with D9 branes and at a
value $T_i$ for the $i^{th}$ complex plane modulus is equivalent to
{\em another} kind of orientifold, with D7-branes and at a value $1/T_i$
of the $i^{th}$ complex plane modulus.

Since the low-energy $\sigma$-model symmetries are seemingly not related
to exact symmetries of the full-string theory, it is not obvious to what
extent these symmetries should be respected at the quantum level. In the
present paper, however, we will argue that the triangle anomalies for these
symmetries may cancel by a GS mechanism in certain specific cases.

This is suggested by the proposed duality \cite{wp}
 between type IIB orientifold
vacua and heterotic compactifications \cite{abpss,kak1,afiv}. For many of
the
models we will study, suitable heterotic orbifold duals have been identified.
Since $\sigma$-model anomalies cancel in these heterotic models, it is
reasonable to expect the corresponding anomalies to cancel also in the
orientifold version.  This does not imply, though, that the anomaly
cancellation pattern is identical. In fact, the detailed analysis of the
triangle anomaly we will perform shows that the anomaly indeed factorizes,
but the GS mechanism required for the cancellation must involve not the
dilaton, but closed string modes in twisted sectors (the dilaton plays a
role only in the cancellation of $\sigma$-gravity mixed anomalies).

Finally, let us remark that we will center on the study of modular
transformations associated to complex planes rotated by all elements in
the orbifold group. Only for such planes we expect the GS mechanism to
cancel the complete anomaly. For complex planes left unrotated by some
element of the orbifold group, both the argument invoking duality with
heterotic models, and the existence of threshold effects in $D=4$
$N=2$ type IIB orientifold \cite{bachas} suggest the anomaly
cancellation may have additional sources beyond the GS mechanism.

\medskip

The paper is organized as follows. In Section~2 we briefly review
$\sigma$-model symmetries in heterotic orbifold vacua. We introduce the
basic notation and discuss the formulae relevant to the cancellation of
anomalies in these models.

In Section~3 we address the same problem in type IIB orientifold vacua. We 
start with a brief review of cancellation of anomalous gauge $U(1)$ 
symmetries, and the generation of Fayet-Iliopoulos terms, in
section~3.1. In section~3.2 we derive general formulae for the triangle
contributions to $\sigma$-model anomalies in type IIB orientifolds, and
compute them explicitly for a set of models. The analysis of the
factorization properties of these anomalies is performed in sections 3.3
and 3.4, where we also discuss how a GS mechanism involving non-trivial
shifts of the RR twisted sector axions may cancel the anomaly. Models with
Wilson lines and/or non coincident branes are studied in section~3.5.
Finally, in section~3.6 we study $\sigma$-gravitational mixed anomalies
and show they can be cancelled through exchange of both the twisted
RR axions and the universal axion (partner of the dilaton).

In Section~4 we comment on the mixing between the untwisted and twisted
moduli required for the GS mechanism to work. An interesting consequence
is the existence of an additional $T_i$-dependent contribution to the FI
terms for the anomalous $U(1)$'s. Finally, Section~5 contains our
conclusions.

\section{Sigma-model anomalies in $D=4$, $N=1$ heterotic orbifolds}

The K\"ahler potential dependence on the complex dilaton $S$ and the
untwisted K\"ahler moduli $T_i$, $i=1,2,3$ in this class of heterotic
orbifolds is well known. These fields live in a coset $\sigma $-model with
symmetry given by \footnote{For particular orbifold models like the $Z_3$
or $Z_6$ there is an enlarged number of untwisted K\"ahler moduli and in
some others like $Z_4$ or $Z_6$' there may be complex structure scalars.
We will concentrate for simplicity on the three K\"ahler moduli $T_i$
which are always present for any orbifold where the six-torus lattice can 
be decomposed as three two-dimensional lattices.}
\beq
{\cal M} \! = \! \Biggl\lbrack{SU(1,1)
\over U(1)}\Biggr\rbrack^3_T \!\! \otimes
\Biggl\lbrack{SU(1,1)
\over U(1)}\Biggr\rbrack_S.
\label{coset}
\eeq
The  K\"ahler potential is  given by
\beq
K(S,S^*,T_i,T_i^*)\ =\ - log(S+S^*)-\sum_{i=1}^3\log(T_i+ T^*_i)
\label{kahler}
\eeq
The kinetic terms for all charged fields $ A_{\alpha  }$ in the orbifold, both
untwisted and twisted may be written to first order in this fields as:
\beq
K^{\rm matter}=\delta_{\alpha\beta}
\prod_{i=1}^3(T_i+ T_i^*)^{n^i_\alpha}
A_\alpha {\bar A}_\beta
\label{matterkin}
\eeq
Here the $n^i_\alpha $, often called modular weights of the fields,  are
constants which depend on the conformal field theory sector corresponding 
to the field. For the untwisted matter fields associated to the $j^{th}$ 
complex plane one finds:
\beq
n_j^i=-\delta_j^i, \  .
\label{modunt}
\eeq
For fields  which are originated from a twisted sector with twist vector
$v=(v_1,v_2,v_3)$ (here we take all $0\leq v_i \leq 1$ and
$\sum_{i=1}^3v_i=1$) one has \footnote{In the presence of twisted
oscillators these formulae are slightly generalized. See ref \cite{il} for
details.}:
\beqa
n_{\alpha }^i\  = & \ -(1\ -\ v_i) \ , & {\rm for}\ v_i \not= 0 
\nonumber\\
n_{\alpha }^i\   = & \  0 \ , & {\rm for}\ v_i \ =\ 0.
\label{modtwist}
\eeqa
The effective classical action presents a $\sigma $-model invariance under
$SL(2,R)_{T_i}$ transformations given by
\beq
T_i\rightarrow{a_iT_i-ib_i\over ic_iT_i+d_i},
\label{dual}
\eeq
with $a_i,b_i,c_i,d_i\in{\bf R}$ and $a_id_i-b_ic_i=1$. Under these
transformations the charged matter fields transform as:
\beq
A_\alpha\rightarrow A_\alpha \prod_{i=1}^3(ic_iT_i+d_i)^{n_\alpha^i}
\label{matert}
\eeq
so that the kinetic terms in (\ref{matterkin}) remain invariant. The
transformation of the superpotential also compensates for the
transformation of the $T_i$-dependent piece in the K\"ahler potential
(\ref{kahler}).

This continuous symmetry is in general expected to be violated by
world-sheet effects. However, in the heterotic case we know that the
discrete subgroup $SL(2,{\bf Z})^3_{T_i}$ of the above non-compact
symmetry corresponds to the T-duality invariance of heterotic vacua. Thus
this discrete subgroup has to remain as a symmetry even after world-sheet
corrections  are included.

In particular, the transformations (\ref{dual}), (\ref{matert}) induce
chiral rotations in the massless fermions of the theory. They are
associated to gauge transformations of a composite gauge vector potential
involving the moduli fields $T_i$. If we compute the triangle anomalies
corresponding to this composite current and two gauge currents one finds
in general an anomalous result. The coefficient of this anomaly can be
computed to be given by \cite{dkl,dfkz,lco} :
\beq
{b'}_a^i=-C(G_a)+\sum_{\underline R_a}
T(\underline R_a)(1+2n_{\underline R_a}^i)
\label{coeff}
\eeq
Here $C(G_a)$ is the quadratic Casimir of the gauge group
$G_a$ in the adjoint representation and $T(\underline R_a)$ is the
quadratic Casimir in the representation ${\underline R}_a$ corresponding
to a charged field. The sum extends over all fields charged under $G_a$
and $n_{\underline R_a}^i$ is the modular weight along the complex plane
$i$ of each given field. In general this mixed $\sigma $-$G_a^2$ anomalies
do not cancel. The gauge kinetic terms get one-loop (non-local) corrections
\cite{dfkz,lco} :
\beq
{\cal L}_{\rm nl}=
\sum_a\int d^2\theta{1\over 4}W^a
W^a\biggl\lbrace S
-
{1\over 32}\partial^{-2}\bar{\cal D}\bar{\cal D}{\cal D\cal D}
 \biggl\lbrack\sum_{i=1}^3{b'}_a^{i}\log(T_i+\bar T_i)
 \biggr\rbrack\biggr\rbrace+{\rm h.c.}
\label{ilu}
\eeq
where $W^a$ are the field strength of gauge fields.
Under $SL(2,{\bf R})_{T_i}$ transformations this action is not
invariant due to the non-local piece.
However we know that discrete $T$-duality transformations have to be a good
symmetry also at the quantum level. The cancellation of the triangle
anomaly comes about from two additional contributions:

{\bf 1)} Under the $\sigma $-model continuous transformations in
(\ref{dual}) the complex dilaton $S$ gets also transformed as
\cite{dfkz,lco,il}
\footnote{We define here $ReS=8\pi^2/g^2$.}:
\beq
S\rightarrow
S+{k_a }\sum_{i=1}^3\delta^i_{GS}\log(ic_iT_i+d_i).
\label{sgssigma}
\eeq
Here $ \delta^i_{GS}$ is a gauge group-independent coefficient which
describes the one-loop mixing between the $S$ and the $T_i$ fields
in these heterotic vacua, and  $k_a$ is the Kac-Moody level of the gauge
group. Since $S$ is the tree-level gauge function coefficient for all
gauge groups, this transformation gives an additional contribution to the
mixed $\sigma $-$G_a^2$ anomalies. In particular this transformation
cancels all  $\sigma $ model anomalies in $Z_N$, $N$ odd orbifolds with no
twist leaving one complex plane unrotated ($Z_3$ and $Z_7$ standard
heterotic orbifolds). This is also the mechanism which cancels anomalies
corresponding to complex planes $i$ which are always rotated by the twists
of the model. Thus, for  example, that is the case of the anomalies with
respect to the first two planes  in the $Z_6$ orbifold generated by the
twist $v=1/6(1,1,-2)$. Notice that this  Green-Schwarz mechanism not only
cancels discrete $T$-duality symmetry anomalies but continuous $\sigma$-model
anomalies. Also notice that this mechanism is gauge group independent and
hence the mixed anomalies should be equal for all gauge groups if they
are to be cancelled only by this mechanism.

{\bf 2) } For heterotic orbifolds containing some complex plane $i$ left
unrotated by some orbifold twist there is a $T_i$-dependent one-loop
threshold correction to the gauge kinetic functions. This threshold
correction is in general gauge-group dependent and was computed in
ref. \cite{dkl}. For complex planes of these type, the discrete
$T$-duality anomalies in (\ref{coeff}) are cancelled by the transformation
properties of these threshold corrections plus the Green-Schwarz
mechanism above. Notice however thus, unlike the previous mechanism, the
threshold correction explicitly violate the continuous $\sigma $ model
symmetries and respect only the discrete subgroups associated to
$T$-dualities.

Notice that in order for the $S$-dependent K\"ahler potential to be
invariant under the transformation in (\ref{sgssigma}) it has to be
modified to:
\beq
K(S,S^*)\ =\ -\log(S+S^* +  {k_a }\sum_i  \delta_{GS}^i \log(T_i+T_i^*)
)
\label{ksgssigma}
\eeq
reflecting explicitly the one-loop mixing between $S$ and $T_i$ fields.

In addition to mixed $\sigma $-gauge anomalies there will be
mixed $\sigma $-gravity anomalies. The corresponding triangle graph
involving massless fermions gives an anomaly proportional to
\cite{il} :
\beq
{b'}_{\rm grav}^i\ =\ 21 \ +\ 1\ +\ \delta _M^i \ -\ {\rm dim}\ G\
+\ \sum _\alpha\ (1\ +\ 2n_\alpha^i)
\label{modgrav}
\eeq
Here 21 comes from the gravitino contribution, $1$ comes from the
dilatino and $\delta _M^i $ represents the contribution from
untwisted moduli. The other two terms come from the contribution
of gauginos and charged chiral matter respectively. The same
two mechanisms which we described above are also present in the
cancellation of the corresponding duality anomalies.

\medskip

Four-dimensional heterotic vacua do also often  have one anomalous gauged
$U(1)$ symmetry in their effective Lagrangian. Those anomalies are
cancelled by a Green-Schwarz mechanism \cite{gs}, very much like the
$\sigma $-model anomalies discussed above. In this case under a gauge
$U(1)$ transformation with gauge parameter $\Lambda (x)$, the dilaton
transforms like:
\beq
S\rightarrow
S+{k_a }\delta_{GS}^X\ \Lambda(x) \ .
\label{sgsu1}
\eeq
and this cancels all mixed anomalies. Notice that in an heterotic model
with both an anomalous $U(1)$ and $\sigma $-model anomalies the
$S$-dependent piece will thus take the form:
\beq
K(S,S^*)\ =\ - \log(S+S^* + {k_a }\sum_i  \delta_{GS}^i \log(T_i+T_i^*)
\ - {k_a }\delta_{GS}^X V_X )
\label{ksgstotal}
\eeq
where $V_X$ is the vector superfield of the anomalous $U(1)$. As is well
known \cite{dsw}, there is also a Fayet-Iliopoulos (FI) term associated to
the anomalous $U(1)$ :
\beq
\xi _{het} \ \propto \  ({ { \partial K}\over {\partial V_X} })_{V_X=0} \ =\
 {  {  {-k_a }\delta_{GS}^X } \over
{ S+S^* + {k_a }\sum_i  \delta_{GS}^i \log(T_i+T_i^*) }  }\
 \label{fihet}
\eeq
This is of the order of the string scale for generic and realistic values
of dilaton and moduli\footnote{The result in eq.(\ref{fihet}) is given
in units of $M_{Planck}^2/8\pi $ since the Kahler potential in
eq.(\ref{ksgstotal}) has in fact such an overall factor.}.

\section{$\sigma$-model anomalies in compact $D=4$, $N=1$ Type IIB
orientifolds}

Let us recall the structure of $D=4$ $N=1$  \cite{bl} -\cite{iru}
type IIB orientifolds \cite
{sagnotti,hor,bs,gp}. The models we will be centering on are constructed
by modding out the toroidally compactified type IIB theory by the joint
action of a symmetry group $G_1$ of the six-torus together with the
world-sheet parity operation $\Omega$ \cite{sagnotti,hor,bs,gp}
. We will consider orientifold
groups  with  the structure $G_1+{\Omega} G_1$ and center on $G_1=Z_N$
orbifold moddings. All possible twists consistent with spacetime $N=1$
supersymmetry have been classified in \cite{dhvw}, and their eigenvalues
$v=(v_1,v_2,v_3)$ are shown in Table~\ref{tzn}.

\begin{table}[htb]
\renewcommand{\arraystretch}{1.25}
\begin{center}
\begin{tabular}{|c|c||c|c||c|c|}
\hline
$Z_3$ & $\frac13(1,1,-2)$ & $Z_6^{\prime}$ & $\frac16(1,2,-3)$ &
$Z_8^{\prime}$ & $\frac18(1,-3,2)$ \\
$Z_4$ & $\frac14(1,1,-2)$ & $Z_7$ & $\frac17(1,2,-3)$ &
$Z_{12}$ & $\frac1{12}(1,-5,4)$ \\
$Z_6$ & $\frac16(1,1,-2)$ & $Z_8$ & $\frac18(1,3,-4)$ &
$Z_{12}^{\prime}$ & $\frac1{12}(1,5,-6)$ \\
\hline
\end{tabular}
\end{center}
\caption{$Z_N$ actions in \Deq4.}
\label{tzn}
\end{table}

The closed string sector is constructed by performing the orientifold
projection to the spectrum of type IIB theory on the corresponding
toroidal orbifold. In the untwisted sector, one obtains the $D=4$ $N=1$
supergravity multiplet, one chiral multiplet $S$ containing the dilaton,
and a further set of moduli describing the geometry of the original
torus. As in the heterotic case mentioned above, the number of such moduli
is model-dependent, and we will center on the three moduli $T_i$,
$i=1,2,3$ corresponding to the three complex planes. The closed  string
twisted modes will also be relevant to our purposes. A fixed point $f$
will have associated chiral singlet fields $M_f^k$ for each $k$-twisted
sector.

The consistency of the equations of motion for the RR potentials requires
the cancellation of the corresponding tadpoles. This is implemented by
introducing D branes whose RR charge cancels that of the orientifold
planes. For $Z_N$, with $N$ odd, only D9-branes are required. They fill
the full space-time and six dimensional compact space. For $N$ even,
D$5_k$-branes, with world-volume filling space-time and the $k^{th}$
complex plane, may be required. This is so whenever the orientifold group
contains the element $\Omega R_i R_j$, for  $k\neq i,j$. Here $R_i$
($R_j$) is an order two twist of the $i^{th}$ ($j^{th}$) complex plane.
In what follows we consider cases with only one set of fivebranes, which,
with the conventions for the twists in Table~\ref{tzn}, wrap the third
complex plane.

The action of an orientifold group element $g$ on D$p$-branes is
specified by a unitary matrix, $\gamma_{g,p}$. It turns out to be useful
to introduce the vectors $V^p$ given in terms of the eigenvalues
\footnote{In order for the orientifold projection to be a symmetry, the
eigenvalues come in complex conjugate pairs. Thus we define $V^p$ to
contain only the phases in $[0,\pi)$.} $e^{2\pi i V_a}$ of the matrices
$\gamma_{\theta,p}$, corresponding to the generators of the orbifold
group. Let us also define $w^p_i$ to be the number of times a given
eigenvalue $V^{p}_i$ appears.

For a given orientifold group one can obtain a set of constraints on the
Chan-Paton matrices for these D branes. Some of them follow from the
requirement that the matrices form a representation of the orientifold
group, while others correspond to cancellation of twisted tadpoles (see
e.g. \cite{afiv} for details). The Chan-Paton matrices determine the
(open-string) spectrum of the model. For instance, the integers $w_i^p$
introduced above specify the ranks of the $i^{th}$ factor in the gauge
group on the D$p$-branes. Instead of entering the details of their
construction, Table 2 conveniently provides a list of the models we will
consider, along with their spectra. The mentioned constraints ensure the
consistency of the resulting models. In particular, they imply the
cancellation of gauge and gravitational anomalies (see \cite{abiu} for a
detailed discussion). It is known that out of the list of twists in
table 1,  the examples based on $Z_4$, $Z_8$, $Z_8'$ and $Z_{12}'$ have
twisted tadpoles and hence are inconsistent \cite{zwart, afiv}, at
least with the standard
GP-projection \cite{gp} here discussed. We will thus focus in the
remaining examples when we treat specific models.

\subsection{Cancellation of $U(1)$ gauge anomalies and FI terms}

Befores studying $\sigma$-model anomalies in type IIB orientifolds, we
recall the structure and cancellation of gauge $U(1)$ anomalies in these
models. This will be useful because there is a close analogy between the
GS anomaly cancellation mechanisms in both cases. Also, the mixing between
untwisted and twisted moduli required to cancel $\sigma$-model anomalies
implies a interesting contribution to the FI terms for the anomalous
$U(1)$'s. Thus this section also provides the notation and formulae
relevant to this issue.

The cancellation of $U(1)$ anomalies in $D=4$ $N=1$ type IIB orientifolds
is quite interesting. In contrast with heterotic models, these
theories have generically several $U(1)$'s with non-vanishing triangle
anomalies. Moreover, the mixed anomalies with different non-abelian gauge
factors and gravity are {\em not} in adequate ratios to be cancelled by a
GS mechanism with exchange of the partner of the dilaton \cite{afiv}.
However, in \cite{iru} it was proposed that a different version of the GS
mechanism, with exchange of RR twisted closed string modes, cancels these
anomalies \footnote{This is the four-dimensional analog of the
six-dimensional mechanism studied in \cite{sagcan,blpssw} . }.

\begin{figure}
\centering
\epsfxsize=4.5in
\hspace*{0in}\vspace*{.2in}
\epsffile{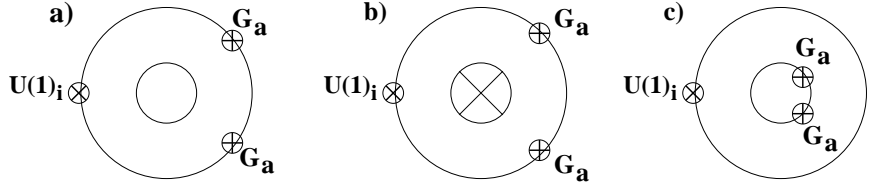}
\caption{\small The string theory diagrams contributing to the three point
amplitude corresponding to the field theory anomaly.
}
\label{diagram}
\end{figure}

\begin{figure}
\centering
\epsfxsize=3.5in
\hspace*{0in}\vspace*{.2in}
\epsffile{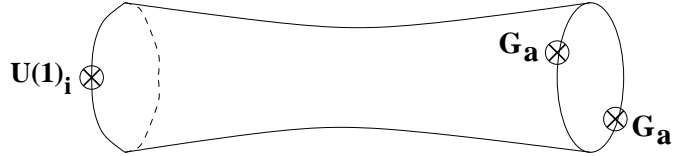}
\caption{\small The string theory diagram in figure 1c in the closed
string channel. This contribution provides the GS counterterms which
cancel the anomaly from the triangle diagrams. Closed string twisted models
propagate in the closed string channel and mediate the GS mechanism.
}
\label{closed}
\end{figure}

Let us consider for instance the mixed non-abelian anomaly. The anomaly in
the field theory is reproduced by the string theory diagrams depicted in
Figure~1, in the point-particle limit in the open-string channel. As
mentioned above, the net contribution is non-vanishing, which would
mean an inconsistency of the theory at the quantum level. However, the
same string diagrams in Figure~1 give additional low-energy contributions
in the point-particle limit in the closed string channel. Notice that the
close-string channel contributions corresponding to diagrams 1a and 1b
cancel against each other, due to cancellation of disk and crosscap pieces
at the end of the infinite tube (this is the statement of tadpole
cancellation). On the other hand, the diagram 1c gives a non-vanishing
closed-string channel contribution, schematically depicted in Figure~2. It
is interpreted as the exchange of twisted RR fields, which propagate on
the infinitely elongated cylinder. This contribution was determined
in\cite{iru} to be
\beq
A_{ij}^{pq}\ = \ {1\over N}\ \ \sum_{k=1}^{N-1} \ C_k^{pq }(v)\
w_i^p\sin2\pi kV^{p }_i \ \cos2\pi kV^{p}_j
\label{masterorient}
\eeq
Here $k$ runs over twisted $Z_N$ sectors, $p$, $q$ run over 5,9 (meaning
5- or 9-brane origin of the gauge boson) and
\beqa
C_k^{pp } & = & \prod _{a=1}^3 2\sin\pi kv_a \quad {\rm for}\;\; p=q
\nonumber\\
 C_k^{59 } & = & 2\sin\pi kv_3
\label{ckpp59}
\eeqa
In \cite{iru} it was checked that this contribution indeed cancels the
mixed non-abelian anomalies in all orientifolds considered. Even though
the structure of the amplitude (\ref{masterorient}) can be read off from
Fig~1c, we would like to stress that the existence of factorization can
be induced from the mere structure of the triangle anomalies. Indeed, a
simple strategy used in \cite{iru}, and to be exploited below, is to write
down the triangle anomaly for a generic orientifold in terms of the
integers $w_i$, which define the Chan-Paton matrices and determine the
field theory spectrum. Then one can perform a discrete Fourier transform
to express the anomaly in terms of the Chan-Paton traces $\Tr \gamma_k$.
At this point, the anomaly exhibits a factorized form with the structure
(\ref{masterorient}). Since consistency of string theory implies the
total contribution from the triangle anomalies plus the GS terms must
vanish, this technique allows to easily obtain the structure of the GS
terms.

We would like to stress that this trick is in the spirit of the study of
many other anomalies in string theory, where a preliminary low-energy
analysis of the anomaly reveals the factorized structure, and suggests
the corresponding GS counterterms. A further step, usually much more
involved, is the explicit string theory computation of the required
couplings.

\begin{figure}
\centering
\epsfxsize=4in
\hspace*{0in}\vspace*{.2in}
\epsffile{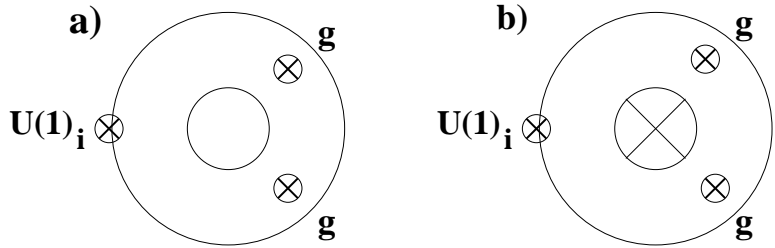}
\caption{\small There are only two string theory diagrams which contribute
to the scattering amplitude of one $U(1)$ gauge boson and two gravitons.
Thus factorization is not as manifest as in the example in figure 1.}
\label{diagrav}
\end{figure}

\begin{figure}
\centering
\epsfxsize=5.5in
\hspace*{0in}\vspace*{.2in}
\epsffile{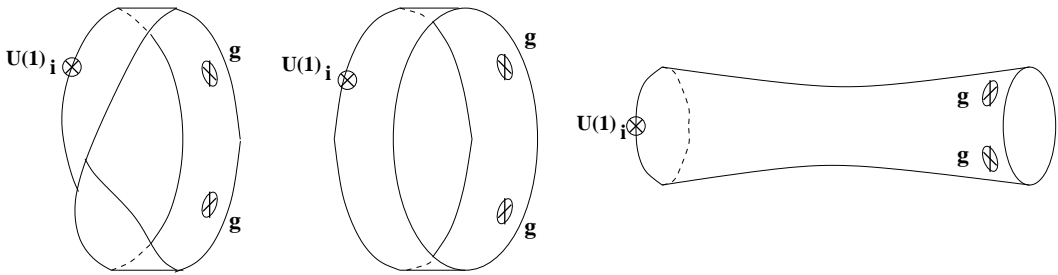}
\caption{\small This figures shows the limits where the string theory
diagrams give field theory contributions. The first two are point-particle
limits in the open string channel and give the usual triangle anomaly. The
third is a point-particle limit in the closed string channel and
represents the exchange of closed string twisted modes. Notice that the
cylinder diagram in figure~\ref{diagrav} generates two different
field-theory contributions.}
\label{closgrav}
\end{figure}

\medskip

The same mechanism can be employed to cancel mixed gravitational
anomalies. Notice that in this case the graviton is a closed string mode,
and so the relevant string diagrammatics is different. In particular, as
shown in Fig~3, only two diagrams contribute and factorization is not
as obvious as in the preceding case. Factorization is, however, strongly
suggested by the structure of the triangle anomalies. This is done as
sketched above: One writes the mixed gravitational anomalies for a general
orientifold, and performs a discrete Fourier transform to rewrite them in
terms of the Chan-Paton matrices. The triangle anomaly is given by a
factorized expression, which can be cancelled by a GS term of the form
\beqa
{A_i^{\alpha}} & = & \frac 34 \frac 1N \sum_{\beta} \sum_{k=1}^{N-1}
C_k^{\alpha \beta}(v)\, w_i \sin \pi k V_i^\alpha (\Tr
(\gamma_k^\beta)^{-1})
\label{irugrav}
\eeqa
In string theory factorization follows from the fact that the cylinder
diagram provides two low-energy contributions, corresponding to the
point-particle limit in the open and closed string channels (fig 4).
Again, Figures~4a and 4b give the usual field theory triangle anomaly,
whereas Figure~4c provides the GS contribution, which can be interpreted
as exchange of twisted RR fields. This type of factorization will be
relevant for the cancellation of $\sigma$ model anomalies that we will
study below.

Before doing that, let us point out some consequences of the coupling of
twisted closed string modes to gauge fields for
the effective low energy field theory Lagrangian.  We will restrict
ourselves in this discussion to $Z_N$ orientifolds with
$N$ odd ($Z_3$ and $Z_7$ in the compact case) since the
structure of their twisted closed string fields is much simpler.
The even $N$ case is discussed in the appendix.
First, for this generalized Green-Schwarz mechanism to
work there must be a modification of the gauge kinetic function.
In particular one has for the gauge group $G_b$:
\beq
 f_b = S + {1\over N}\sum_{k=1}^{(N-1)/2} \cos(2\pi kV_b) {1\over C_k}\sum_f M_f^k
\label{ffunc}
\eeq
Here the sum on $f$ goes over the number of fixed points $=C_k^2$
whereas $k$ labels the twisted sectors. $M_f^k$ is a closed string
chiral singlet field living on the fixed point $f$ and corresponding to
the twist $\theta ^k$. Notice that, since the D9-brane world-volume spans
the complete compact space, the gauge fields couple to twisted fields from
all fixed points. Now, under a $U(1)_a$ gauge transformation with
parameter $\Lambda_a(x)$ the twisted closed string chiral fields $M_i^k$
transform as:
\beq
Im M_f^k \rightarrow  Im M_f^k\ + \ w_a\, 2\, \sin(2\pi kV_a) \ \Lambda_a(x)
\label{u1shift}
\eeq
Notice how in this way the net effect of this shift combined
with eq.(\ref{ffunc}) is the contribution (\ref{masterorient}) discussed
above.

The fact that the fields $M_f^k$ are not gauge invariant means that the
K\"ahler potential of those fields must have the general dependence:
\beq
K(M_f^k,{M_f^k}^*)\ =\ K(M_f^k+{M_f^k}^* \ - \sum_a \delta_{GS \, k}^a V_a )
\label{kaeme}
\eeq
where the sum on $a$ goes over the $U(1)$'s in the model and
\beq
\delta_{GS \, k}^a\ =\  w_a\, 2\, sin(2\pi k V_a)
\label{dgsm}
\eeq
Among the interactions generated upon expansion in components, a
particularly interesting piece is a FI term for $U(1)_a$. If, as pointed
out in ref.\cite{poppitz} the K\"ahler potential for the fields $M_k$ is
bilinear \footnote{The same qualitative conclusion follows for more
general K\"ahler potentials, as long as they are non-singular at the
orientifold point \cite{nilles}. We would like to thank E.~Poppitz for
comments on this point.}, given by $(M_k^i+{M_k^i}^*)^2$, the FI term is 
\beq \xi_{IIB}^a \ = \ -
\sum_f \sum_k  { w_a\, 2\, sin(2\pi k V_a) } ( M_f^k+{M_f^k}^*)
 \label{fiIIB}
\eeq
In models with 5-branes the results are analogous, and may be obtained
starting from the results in the appendix.

A general property of this contributions to the FI-term is that it is
controlled by the blow-up modes $Re M_f^k$, and can be adjusted at will.
In particular, this contribution vanishes at the orbifold point. The above
result concerning FI-terms will be revised in Section 4 in the light of
our analysis of $\sigma$-model anomalies, since it implies substantial
changes in several respects.

\medskip

\subsection{ $\sigma $-model anomalies }

Let us turn to the study of $\sigma$-model anomalies in $D=4$ $N=1$ compact
orientifolds. As mentioned in the introduction, heterotic/type I duality
suggests that these anomalies should also  cancel at least in some classes
of  type IIB orientifold. More precisely, there are a number of $D=4$,
$N=1$ orientifolds for which specific candidate heterotic duals have been
identified (see e.g. \cite{afiv} for a general discussion). That is the
case for example of the $Z_3$ \cite{abpss,kak1} and  $Z_7$ \cite{kak2}
orientifolds. Now, we already mentioned in the previous chapter that
$\sigma$-model anomalies in the heterotic side are only cancelled for
complex planes which are rotated by all twists in the orbifold group.
Thus, properly speaking, we should expect cancellation of $\sigma$-model
anomalies in the type IIB orientifold case only for this type of
complex direction. Thus, they should cancel for any complex plane in $Z_N$
orientifolds with $N$ odd, and also along some complex planes of other
type of orientifolds, like e.g., the first two complex planes of the $Z_6$
and $Z_{12}$
orientifolds  or the first complex plane in the $Z_6'$
orientifold. We will concentrate our study to those complex planes in
which the candidate heterotic duals present cancellation of $\sigma$-model
anomalies.

\begin{figure}
\centering
\epsfxsize=4in
\hspace*{0in}\vspace*{.2in}
\epsffile{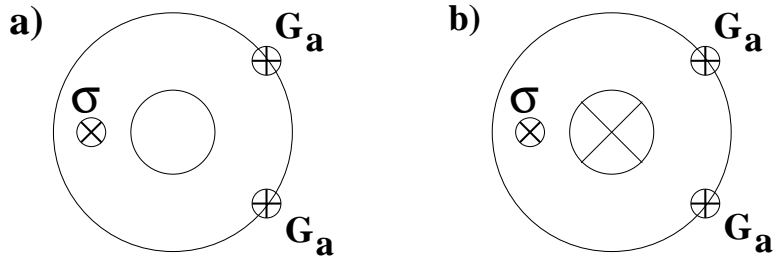}
\caption{\small The two string theory diagrams contributing to the
coupling of the $\sigma$-model composite connection with two non-abelian
gauge bosons.}
\label{diagmod}
\end{figure}

\begin{figure}
\centering
\epsfxsize=5.5in
\hspace*{0in}\vspace*{.2in}
\epsffile{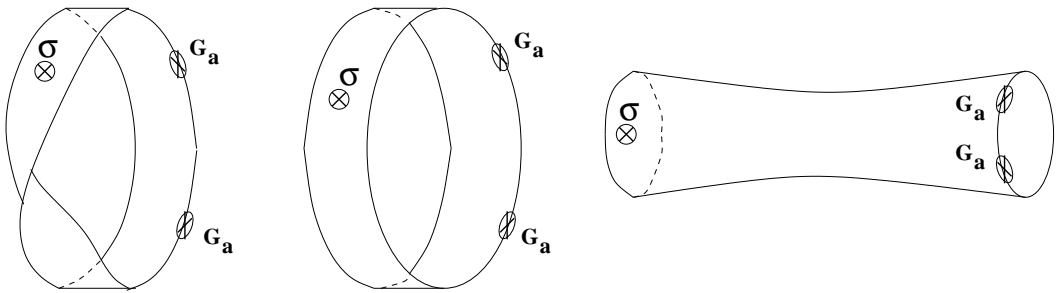}
\caption{\small Factorization in this case is expected to follow the same
pattern encountered for gravitational-U(1) gauge mixed anomalies.}
\label{closmod}
\end{figure}

It is easy to compute directly the triangle anomalies in several models,
and check that they do not cancel. Let us consider the simplest Type IIB
$D=4$, $N=1$ orientifolds $Z_3$, $Z_7$, $Z_6$, $Z_6$' and $Z_{12}$ whose
spectra are given in Table~\ref{specs} for a configuration with all 5-branes
sitting at the origin (see ref.\cite{afiv} for details). In order to
compute the anomalies we need to know  the ``modular weights'' $n_i$ of
each field with respect to $SL(2,{\bf R})_{T_i}$ transformations. In
other words, we need to know the $T_i$ dependence of the kinetic terms of
each field in the orientifold. For this class of orientifolds this was
discussed in ref.\cite{afiv}  and \cite{imr}. For models with only
9-branes the relevant piece of the K\"ahler potential is analogous to that
of the untwisted sector of the heterotic orbifolds, namely
\beq
K(S,S^*,T_i,T_i^*)\ =\ - \log(S+S^*)-\sum_{i=1}^3\log(T_i+ T^*_i 
-|C_i^9|^2)
\label{kahlerunt}
\eeq
where $C_i^9$ are the charged fields from the $(99)$ sector of the orbifold
corresponding to the $i^{th}$ complex plane. Thus the ``modular weights'' of
the $C_i^9$ field with respect to the $j^{th}$ complex plane is
$n_i^j=-\delta_i^j$. For models with 5-branes, like the $Z_6$, $Z_6$' and
$Z_{12}$  in Table~\ref{specs}, the corresponding K\"ahler potential is
given by:
\beqa
K &=& -\log(S+S^* + |C_3^5|^2)   -  \log(T_3+T_3^* + |C_3^9|^2 )
\nonumber \\[0.2ex]
   & - & \log(T_1+T_1^*+|C_1^9|^2 +|C_2^5|^2 )  -
\log(T_2+T_2^*+|C_2^9|^2 +|C_1^5|^2 )
\nonumber\\[0.2ex]
  &+ & {{|C^{95}|^2}\over {(T_1+T_1^*)^{1/2}(T_2+T_2^*)^{1/2}}}
\label{kali}
\eeqa
Here the world-volume of 5-branes includes the third complex plane. Also,
$C^{95}$ are the charged fields in the $(95)$ sector of the orientifold.
The $S$ and $T_i$ dependence of the gauge kinetic functions is given by
$f_9=S$ and $f_5=T_3$. Notice that, since the gauge kinetic function for
the 5-brane gauge group is $T_3$, the $(55)$ gauge group explicitly
breaks the $SL(2,{\bf R})$ symmetry associated to $T_3$ and hence
cancellation of  $\sigma $-model anomalies in the third complex plane
are not in principle expected.

>From the above formula one sees that the ``modular weights''
along the $j^{th}$ plane ($j=1,2$) are given by:
\beq
n^i_9\ =\ -\delta_i^j \ ; \  n^i_5\ =\ \delta_i^j-1 \ ;
n^1_{95}\ =\ -1/2  \ ; n^2_{95}\ = \ -1/2 \ ; n^3_{95}\ =\ 0  \ .
\label{pesos}
\eeq
%


\begin{table}[pht!]
\footnotesize
\renewcommand{\arraystretch}{1.25}
\begin{center}
\begin{tabular}{|c|c|c|}
\hline
Twist Group   & & \\
\cline{1-1}
Gauge Group & \raisebox{2.5ex}[0cm][0cm]{ (99)/(55) matter} &
\raisebox{2.5ex}[0cm][0cm]{ (95) matter}  \\
\hline\hline
$Z_3 $ & $(8,12)_1 + (1,{\overline {66}})_1 $  &  -  \\
\cline{1-1}
$SO(8)\times U(12)$ &  $(8,12)_2 + (1,{\overline {66}})_2$  & \\
                    &   $(8,12)_3 + (1,{\overline {66}})_3$ & \\
\hline\hline
$Z_7$
& $ (8,{\ov 4},1,1)_1 + (1,4,{\ov 4},1)_1 + (1,1,4,{\ov 4})_1+(1,1,1,6)_1  $
&  -  \\
\cline{1-1}
$SO(8) \times U(4)^3$
& $(8,1,{\ov 4},1)_2 + (1,4,1,{\ov 4})_2 + (1,1,4,4)_2 + (1,6,1,1)_2$ & \\
 &  $(8,1,1,4)_3 + (1,{\ov 4},1,{\ov 4})_3 + (1,4,4,1)_3 + (1,1,6,1)_3$
& \\
\hline\hline
$Z_6$
& $(15, 1, 1)_1 + ({\ov 6},4,1)_1 + (1,{\ov 4},6)_1+(1,1,{\ov{15}})_1 $
&  $(6,1,1;6,1,1) + (1,1,{\ov 6};1,1,{\ov 6})$ \\
\cline{1-1}
$[ U(6)\times U(4)\times U(6) ]^2$
& $(15, 1, 1)_2 + ({\ov 6},4,1)_2 + (1,{\ov 4},6)_2+(1,1,{\ov{15}})_2$
& $(1,1,6;1,\ov{4},1) +(1,\ov{4},1;1,1,6)+ $  \\
& $({\ov 6},{\ov 4},1)_3 + (6,1,{\ov 6})_3 + (1,4,6)_3 $
& $({\overline 6},1, 1; 1,4,1)+ (1,4,1; {\ov 6},1,1)$ \\
\hline\hline
$Z_6'$
& $(6,1,1)_1 + ({\ov 4},8,1)_1 + (1,{\ov 8},4)_1 + (1,1,6)_1 $
& $({\overline 4},1,1;{\overline 4},1,1)+(1,1,4;1,1,4)+$ \\
\cline{1-1}
$[U(4)\times U(8)\times U(4)]^2 $
& $(4,8,1)_2 + ({\ov 4},1,4)_2 + (1,{\ov 8},{\ov 4})_2$
& $(1,1,{\overline 4};1,8,1)+  (1,8,1;1,1,{\overline 4})+$ \\
& $(1,28,1)_3 + (1,{\overline {28}},1)_3 + (4,1,4)_3 + ({\ov 4},1,{\ov
4})_3 $
& $(4,1,1;1,{\overline 8},1) + (1,{\overline 8},1; 4,1,1)$ \\
\hline\hline
$Z_{12}  $ 
& $ (1,{\ov 3},{\ov 3},1,1,1)_1 + ({\underline{3,1}},1,1,2,1)_1 + 
(1,1,{\underline{3,1},1,2})_1 $ 
&  $({\ov 3},1^5;1,{\ov 3},1^4) + (1,3, 1^4; 1^5,2)+$   \\
\cline{1-1}
&  $ (3,1,1,1,1,1)_1 + (1,1,1,3,1,1)_1 + ({\ov 3},1,1,{\ov 3},1,1)_2 $
&      $ (3,1^5;1^4,2,1)+ (1^2,3,1^3;1^4,2,1) + $  \\
$(U(3)^4\times U(2)^2 )^2$   
& $(1,1,{\underline{3,1}},2,1)_2 + ({\underline{3,1}},1,1,1,2)_2 +
(1,{\underline{3,1}},1,1,1)_2 $ 
& $(1^2,{\ov 3},1^3;1^3,{\ov 3},1^2) + (1^3,3,1^2;1^5,2)$ \\
& $ ({\ov 3},1,{\ov 3},1,1,1)_3 + (1,{\ov 3},1,{\ov 3},1,1)_3 + 
(3,1,1,1,1,2)_3$ 
&  + same with groups reversed  \\
& $ (1,1,3,1,1,2)_3 + (1,1,1,3,2,1)_3 + (1,3,1,1,2,1)_3 $ & \\
\hline\hline
\end{tabular}
\end{center}
\caption{
Gauge group and charged chiral multiplets in some $Z_N$ ,
\Deq4, \Neq1 type IIB orientifolds. The subindices denote the complex 
planes associated to the different matter fields. Underlining is used to 
indicate the spectrum contains the all permutations of the underlined 
representations.}
\label{specs}
\end{table}

The mixed K\"ahler-gauge anomalies with respect to the
three complex planes can be computed now using eq.(\ref{coeff}).
The results are as follows:

${\bf Z_3 : }$
The anomalies with respect to $SU(12)$ and $SO(8)$ are:
\beqa
({b^{i}}'_a) \ = \ \bmat{cc}
-3 & 6 \\
-3 & 6 \\
-3 & 6
\emat
\label{anomz3}
\eeqa
${\bf Z_7:}$
The mixed anomalies with respect to the $SU(4)^3\times SO(8)$ yield:
\begin{equation}
 ({b^i}'_a)\ =\   \bmat{cccc}
-3  & 1 & 3 & -2 \\
3 &  -3   & 1 & -2\\
1 & 3 & -3   & -2
\emat
\label{anomz7}
\end{equation}
${\bf Z_6: }$
The mixed anomalies with respect to the 9-brane group
$SU(6)^2\times SU(4)$ are:
\begin{equation}
({b^i}'_a)\ =\   \bmat{ccc}
-1  & -1  & 2 \\
-1 &  -1   & 2\\
  2  & 2   & 8
\emat
\label{anomz6}
\end{equation}
${\bf Z_6': }$
The mixed anomalies with respect to the 9-brane group
$SU(4)^2\times SU(8)$ are:
\begin{equation}
({b^i}'_a)\ =\   \bmat{ccc}
3  & 3  & -6 \\
1 &  1   &-2\\
  5  & 5   & 2
\emat
\label{anomz6prima}
\end{equation}
${\bf Z_{12}: }$
The mixed anomalies with respect to the 9-brane group
$SU(3)^4\times SU(2)^2$ are:
\begin{equation}
({b^i}'_a)\ =\   \bmat{cccccc}
1/2  & -3/2  & -3/2 & 1/2 & 1 & 1 \\
-3/2 &  1/2  & 1/2 & -3/2 & 1 & 1 \\
  1  &  1 & 1 & 1 & 4 & 4
\emat
\label{anomz12prima}
\end{equation}
In principle one expects cancellation of $\sigma $-model anomalies with
respect to the three complex planes for $Z_3$, $Z_7$, with respect to the
first two complex planes for $Z_6$ and $Z_{12}$ and only with respect
to the first one for $Z_6$'. Notice that the mixed anomalies with
different gauge factors  are {\em not} in appropriate ratios to be
cancelled by a GS mechanism with exchange of the partner of the dilaton
as in heterotic vacua.

This is hardly surprising, since we are already familiar with the fact
that type IIB orientifolds GS interactions are typically mediated
by exchange or RR fields in the closed string twisted sector. In the
following we will argue that these interactions provide a mechanism to
cancel $\sigma$-model anomalies.

In the particular case of mixed non-abelian anomalies, the relevant string
diagrams, shown in figure~\ref{diagmod} are analogous to those appearing
in the study of gravitational- gauge $U(1)$ mixed anomalies, reviewed
above. In
particular, only two topologies contribute to the string theory
amplitude. This is so because  the composite gauge connections associated
to the $\sigma$-model symmetries are constructed out of closed string
moduli.
Thus, we expect a factorization pattern (see  see fig~\ref{closmod})
analogous to that found in gravitational- gauge $U(1)$ anomalies.

Further support can be obtained through a detailed study of the triangle
anomalies. As explained above, a simple technique to detect factorization
is to write the general triangle anomaly and then perform a Fourier
transform as we  show in the next section.

\medskip

An important final point we would like to remark is the relationship
between the $\sigma $-model anomalies and the conformal anomaly in this
type of orientifolds. If we consider the diagonal $SL(2,{\bf R})_T$
transformation corresponding to the identification $T_1=T_2=T_3=T$, the
corresponding anomaly coefficient is given by:
\beq
b^T_a\ =\ \sum_{i=1}^3 \ {b^i_a}' \ =\ \beta _a
\label{beta}
\eeq
i.e., it equals the one-loop $\beta_a$ function. This happens because
(unlike the heterotic case) all charged fields have  overall ``modular
weights'' $n^T=-1$. Thus for all conformal theories (like, e.g. subsectors
of the theory corresponding to branes away from orientifold planes) the
overall anomalies cancel identically.

\subsection{Green-Schwarz cancellation of $\sigma $-model anomalies:
 odd $N$, $Z_N$ orientifolds}

The only odd order $Z_N$ twists which act cristalographically in six
compact dimensions are $Z_3$ and $Z_7$, but we will write general
expressions for $Z_N$ with $N=2P+1$ with arbitrary $P$. The gauge group in
this class of orientifolds is given by
\begin{equation}
SO(w_0)\times \prod _{j=1}^P U(w_j)
\label{gg9}
\end{equation}
and the charged chiral fields from the $(99)$ sector are given by:
\beq
\sum_{i=1}^3\ (\sum _{a=0}^{2P}\ {1\over 2} ({\mbox {\boldmath $w_a$}},
 {\mbox {\boldmath $w_{a+l_i}$}})\
+ {\bf A_{{N-l_i}\over 2}} + {\bf A_{{-l_i}\over 2}} \ )
\label{specodd}
\eeq
where $v_i=l_i/N$, and {\boldmath $w$}, ${\bf A}$ denote the fundamental and
two-index antisymmetric representations. The sum over $a$ goes only over
$2a\not=l_i$, $mod \ N$ and a negative subindex for a representation implies
conjugation. For fractional subindices the corresponding antisymmetric
representations are absent. Starting from this spectrum and using
eq.(\ref{coeff}) one obtains the following $\sigma $-$G_a^2$ anomalies
with respect to the $i^{th}$ complex plane:
\beqa
{b^i}\, '_a\ & =& \ -w_a+2\delta_{a,0}-{1\over 2} (w_{a+l_i}+w_{a-l_i})
+\delta_{2a+l_i}+\delta_{2a-l_i}\\
&+& \ {1\over 2} \sum_{j\not=i}(w_{a+l_j}+w_{a-l_j})- \sum_{j\not=i}
(\delta_{2a+l_j}+\delta_{2a-l_j})
\label{anozodd}
\eeqa
Now we want to re-express this in terms of the Chan-Paton twist matrices.
The general action of the twists $k=1,\ldots,N$  on the 9-branes is given
by the matrix:
\beq
{ \gamma }_k   = \diag  (I_{w_0}, \alpha_k I_{w_1},
\cdots, \alpha_k^{j} I_{w_j},\cdots, \alpha_k^{N-1} I_{w_{N-1}})
\label{g9}
\eeq
with $\alpha_k = {\rm e}^{2i\pi k/N}$. Notice that the orientifold
symmetry requires $w_a=w_{N-a}$. The trace of this matrix is given by
\beq
\Tr\gamma_k\ =\ \sum_{a=1}^N e^{2\pi i k\over N} w_a
\label{traza}
\eeq
Now we can perform an inverse Fourier transform to re-express the $w_a$ in
terms of the traces of $\gamma _k$'s. Plugging it back in eq.(\ref{anozodd})
one obtains, after some simple algebra, the result:
\beq
{b^i}\, '_a\ = \ {1\over {N} } \sum_{k=1}^{N-1}  {\tilde {\alpha}}_{ k}^i
\cos(4\pi { k}{\bar V}_a)
\label{anomotra}
\eeq
with
\beq
{\tilde {\alpha}}_{ k}^i ={1\over 2} C_{2k}(v) \cotg(2\pi{k}{v}_i)
\Tr\gamma_{2k} -   C_k(v) \cotg(\pi{k}{v}_i)
\label{alphaodd}
\eeq
where $C_k$ was defined in eq.(\ref{ckpp59}). Notice that this
expression has the structure expected from the diagrams in Fig.5.
Indeed, the factor $\cos(4\pi { k}{ V}_a) = \Tr(\gamma_{2k}\lambda_a^2)$
comes from the insertion of two gauge bosons in the outer boundaries of
figures 5a and 5b. Then the cylinder graph should give rise to the
$\Tr\gamma_{2k}$ factor in eq.(\ref{alphaodd}) whereas the other
term corresponds to the Moebius strip graph 5a. It is nice to recover
the expected structure starting merely from the massless spectrum.
Twisted tadpole constraints in odd $N$ orientifolds further require
\cite{abpss,kak1,zwart,afiv} :
\beq
\Tr \gamma _{2{ k}}\ =\ 32 \prod_{i=1}^3 \cos(\pi k v_i)
\label{tadpodd}
\eeq
and plugging this back into eq.({\ref{alphaodd}) one finally gets
the simple result:
\beq
{\tilde {\alpha}}_{ k}^i =-C_{k}(v) \;\tg(\pi{ k}v_i)
\label{tangente}
\eeq
where we have made use of the fact that
$C_{2k}=C_{k}\prod_i 2\cos(\pi kv_i) $. Thus altogether the mixed
$\sigma$-gauge anomalies for odd order orientifolds can be written as:
\beq
{b^i}\, '_a\ = \ {{-2}\over N}\ \ \sum_{k=1}^{(N-1)/2} \ C_k(v)\
tg(\pi kv_i)  \ \cos4\pi kV_a
\label{masterorientm}
\eeq
A comparison with the equivalent result for mixed gauged $U(1)$ anomalies
in eq.(\ref{masterorient}) shows how analogous these expressions are.
This is highly suggestive that indeed, as it happened in the $U(1)$ case,
$\sigma$-model anomalies are cancelled by a Green-Schwarz mechanism in which
twisted RR fields are exchanged in the closed string channel. From the
field theory point of view, the mechanism works in analogy with the
discussion following eqs (\ref{ffunc}) and (\ref{u1shift}). In the present
case the twisted fields corresponding to the fixed point $f$ would
transform with respect to a $\sigma $-model transformation along the
$i^{th}$ complex plane like:
\beq
Im M_f^k \rightarrow  Im M_f^k\ + \ {2tg(\pi k v_i)} \
\log(ic_iT_i+d_i)
\label{shiftm}
\eeq
which combined with (\ref{ffunc}) would exactly cancel the anomaly
(recall $C_k^2$ gives the number of fixed points in these compact
orbifols).

Let us finally comment that using eqs.(\ref{beta}) and (\ref{masterorientm})
one can write a simple expression for the $\beta $-functions of the gauge
groups in these models:
\beq
\beta_a = - {2\over N}  \ \sum_{k=1}^{(N-1)/2}\ {{C_k^2}\over
{ \prod_{i=1}^32 \cos(\pi k v_i)}}  \cos(4\pi k V_a)=
 -{2\over N} \sum_f \sum_k  {1 \over
{ \prod_{i=1}^3 2 \cos(\pi k v_i)}}  \cos(4\pi k V_a)
\label{ellabeta}
\eeq
where $f$ label the fixed points.

For the compact $Z_3$ and $Z_7$ orientifolds under consideration, if one
sets all $M_f^k=M$ the gauge kinetic function (\ref{ffunc}) may be written
in the simple form
\beq
f_b \ =\ S \  \pm   \ { {\beta _a}\over 2}\ M
\label{cachonda}
\eeq
for $Z_3$ and $Z_7$ respectively. Under an $SL(2,{\bf R})_T$
transformation one has $ImM\rightarrow ImM\pm 2\log(icT+d)$, for $Z_7$ and
$Z_3$ respectively and this cancels the overall modulus anomaly. Indeed the
variation cancels the contribution from the second term in (\ref{ilu}).
This is an interesting expression since the disk coupling of the
twisted field $M$ looks like a one-loop factor, in the
sense that it is proportional to the $\beta $-function. This fact had
already been observed in ref.\cite{imr}, where it was used to achieve
precocious unification of gauge couplings constants.

\subsection{Green-Schwarz cancellation of $\sigma$-model anomalies:
 even  $N$, $Z_N$ orientifolds}

Here we will discuss the particular case of even order $Z_N$ orientifolds
with only one sector of 5-branes, with world-volume in the third
complex direction. As we mentioned above, in this case $SL(2,{\bf R})_{T_3}$
is explicitly broken at the classical level  by the gauge kinetic terms
of the 5-brane gauge group so we will not discuss  anomaly cancellations
along the third complex plane. In addition, if the complex plane is left
unrotated in some twisted sector, we know that in the heterotic dual
anomalies are not cancelled only by a GS mechanism. Thus  we will discus
anomaly cancellation only along complex planes which are rotated by all
twists in the model.

We will consider the case of the standard Gimon-Polchinski projection
\cite{gp} leading to a embedding ``without vector structure'' in the gauge
degrees of freedom. The prototype models we have in mind here are the
$Z_6$ and $Z_6$' orientifolds mentioned above.  Let us  consider arbitrary
$Z_N$ ($N=2P$) twists with eigenvalues given by $\frac{1}N (l_1,l_2,l_3)$,
with $l_1+l_2+l_3=0$  and $l_3$ an even integer (thus $l_1,l_2$ are odd).
As we said, we  concentrate on models without vector structure. The
general Chan-Paton matrix for D9-branes has the form
{\small
\beq
{ \gamma_{k,9} }   =
\diag  (\alpha_k I_{w_1},\cdots,\alpha_k^{(2j-1)}I_{w_j},\cdots,
\alpha_k^{(2P-1)} I_{w_P}, \alpha_k^{-(2P-1)} I_{w_P},\cdots,
\alpha_k^{-(2j-1)} I_{w_j},\cdots,\alpha_k^{-1} I_{w_1})
\label{g5}
\eeq
}
with $\alpha_k = {\rm e}^{i\pi k /N}$. Here we have already imposed the
orientifold symmetry $w_{j}=w_{N-j+1}$. The matrices for D5-branes are
analogous with the replacement of $w_j$ by $u_j$ as the number of
eigenvalues $\alpha_k^{(2j-1)}$. These matrices correspond to the shifts
\begin{equation}
V^{p }={\frac{1}{2N}} (1,\cdots 1,\cdots,2j-1,\cdots,2j-1,\cdots,2P-1,
\cdots, 2P-1)
\label{vp}
\end{equation}
with $w_j$ ($u_j$) entries $(2j-1)$ for D9- and D5-branes, respectively,
and $j=1,\cdots,P$.

The associated  gauge group is
\begin{equation}
\prod _{j=1}^{P} U(w_j)\times \prod _{j=1}^{P} U(u_j)
\label{gg5}
\eeq
The complete massless spectrum for this class of models can be found in
ref.\cite{abiu}. Using that spectrum and the modular weights given in
eq.(\ref{pesos}), one finds the following result for the mixed anomalies
with the non-Abelian gauge symmetries from the 9-brane sector:
\beqa
{b^i}\, '_a\ & =& \ -w_a - {1\over 2} (w_{a+l_i}+w_{a-l_i})
+\delta_{2a+l_i+1}+\delta_{2a-l_i+1}\\ \nonumber
&+& \ {1\over 2} \sum_{j\not=i}(w_{a+l_j}+w_{a-l_j})- \sum_{j\not=i}
(\delta_{2a+l_j-1}+\delta_{2a-l_j-1})
\label{anozeven}
\eeqa
The trace of the twist matrices are given by
\beq
\Tr\gamma_{k,9}\ =\ \sum_{a=1}^N e^{{i\pi (2a-1) k}\over N} w_a
\label{trazaeven}
\eeq
(and analogously for 5-branes). Again, we perform an inverse discrete
Fourier transform to express the $w_a$ in terms of the Chan-Paton
traces and substitute them in (\ref{anozeven}). The result is
\beq
{b^i}'_a\ = \ {1\over {2N} } \sum_{k=0}^{N-1}  {\tilde {\alpha}}_{ k}^i
\cos(4\pi { k}{\bar V}_a)
\label{anom}
\eeq
where
\beqa
{\tilde {\alpha }}_{ k}^i\ &  =& \
(\sum_{j\not=i}\cos(4\pi kv_j)  - \cos(4\pi kv_i)-1)\ \Tr\gamma_{2k,9}
\nonumber\\
& -& 4(\sum_{j\not=i}\cos(2\pi kv_j) - \cos(2\pi kv_i)) \ +\ \delta_3^i\
\cos(2\pi kv_3)\ \Tr\gamma_{2k,5}
\label{imporpar}
\eeqa
After some trigonometry  and rearrangement of terms 
one can rewrite this formula
as:
\beq
{\tilde {\alpha}}_{ k}^i ={1\over 2} C_{2k}(v)\ \cotg(2\pi{ k}{v}_i)
\Tr\gamma_{2k_9}
  -  2\ C_k(v)\ \cotg(\pi{ k}{ v}_i)\ +\ \delta_3^i\ \cos(2\pi kv_3)
\ \Tr\gamma_{2k,5}
\label{alphaevenn}
\eeq
where  the sum in eq.(\ref{anom}) is now extended only from
$k=1$ to $k=N-1$.
Now, if we restrict to the case of complex planes $i$ which are rotated
by all twists in the model, the last term drops, leaving
\beq
{\tilde {\alpha }}_{ k}^i\   = \
{1\over 2}\ C_{2k}\ \cotg(2\pi kv_i)\ { {\Tr\gamma_{2k,9}}}\ -\ 2\ C_k\
\cotg(\pi kv_i)
\label{imporporuno}
\eeq
To proceed further we need to impose the twisted tadpole cancellation
conditions for $\Tr\gamma_{2k,9}$, which are model dependent. For the
$Z_6$ and $Z_6$' orientifolds we have the condition
\footnote{The difference in sign compared to eq.(\ref{tadpodd}) is due to the
fact that here $\gamma_{1,9}^N=-1$ since the embedding has no vector
structure.}:
\beq
\Tr \gamma _{2k,9}\ =\ (-1)^k\ 32\ \prod_{i=1}^3 \cos(\pi k v_i)
\eeq
After substitution one finally gets:
\beq
{\tilde {\alpha }}_{ k}^i\   = \
2\ [ \ (-1)^k\ C_{4k}\ \cotg(2\pi kv_i) \ -\ C_k\ \cotg(\pi kv_i)\ ]
\label{imporpordos}
\eeq
The results for $Z_6$ and $Z_6$' for the different twists $k$
are shown in Table~\ref{betazn}.

\medskip

For the $Z_{12}$ orientifold with all the D5-branes at the origin we have
the following twisted tadpole cancellation conditions \cite{afiv,abiu} :
\beq
\Tr \g_{k,9}  = \Tr \g_{k,5} = 0 \quad ; \quad k=1,2,3,5,7,9,10,11
\label{z12a}
\eeq
\beqa
\Tr \g_{4,9} & = &  \Tr \g_{4,5} \ \, = \ \, 4
\nonumber \\[0.2ex]
\Tr \g_{8,9} & = &  \Tr \g_{8,5} \ \,  = \ \,-4
\nonumber \\[0.2ex]
\label{z12b}
\eeqa

Plugging these traces back into eq.({\ref{imporporuno}), one finally gets the
${\tilde {\alpha }}_k^i$. For the $k$=odd contribution the first term in
eq.({\ref{imporporuno}) vanishes,
leaving only:
\beq
{\tilde{\alpha}}_{k}^i\   = \ \ -2C_k\ \cotg(\pi kv_i) \ , \  i=1,2 \ ,\
k={\rm odd}
\label{z12c}
\eeq
For the k=even sectors the contribution of $\Tr \g_{2k}$ is non-vanishing, The
additional contribution for these sectors is of the form:
\beq
\ \pm 2C_{2k}\ \cotg(2\pi kv_i) \ , \  i=1,2 \ ,\ k={\rm even}
\label{z12d}
\eeq
The sum of the two terms gives the ${\tilde {\alpha }}_k^i$ shown in
Table~\ref{betazn}.

\begin{table}[htb]
\renewcommand{\arraystretch}{1.25}
\begin{center}
\begin{tabular}{|c||c|c|}
\hline
  & ${\tilde {\alpha }}_k^i$  &  $ i, k$  \\
\hline
$Z_3$,$Z_7$  & $ -C_k\ \tg(\pi kv_i)$  & $i=1,2,3$ \  ,\  $k=1,..,(N-1)/2$ \\
\hline

$Z_6$ &  0 & $i=1,2$ \ ,\ $k=1,3,5$  \\
\hline
      & $ -C_k\ (\tg(\pi kv_i)+\cotg(\pi kv_i))$  & $i=1,2$\ ,\  $k=2,4$ \\
\hline
$Z_6$' &  $ -2C_k\ \cotg(\pi kv_i)$   & $i=1$\ ,\ $k=1,5$ \\
\hline
      &   0   &  $i=1$\ , \ $k=2,3,4$\\
\hline
$Z_{12}$ &  $-2C_k\ \cotg(\pi kv_i)$ & $i=1,2$ \ ,\ $k=$odd  \\
\hline
      & $ 2\ (C_{2k}\ \cotg(2 \pi kv_i)-C_k\ \cotg(\pi kv_i))$  &
$i=1,2$\ ,\  $k= 2$ mod 4  \\
\hline
      & $ -2\ (C_{2k}\ \cotg(2 \pi kv_i)+C_k\ \cotg(\pi kv_i))$  &
$i=1,2$\ ,\  $k=0$ mod 4 \\

\hline
\end{tabular}
\end{center}
\caption{${\tilde {\alpha }}_k^i$ coefficients for some orientifolds.}
\label{betazn}
\end{table}

As we see, the structure we obtain for the $\sigma $ model anomalies along
complex planes which are always rotated by the twists is very analogous to
that we found for odd orientifolds. The result in both cases shows the
anomaly can be cancelled by a GS mechanism mediated by the exchange of
twisted RR fields. Comparing these results with those obtained for
gauged anomalous $U(1)$'s, one observes  that the role of the
$\sin(2\pi kV_a)$ factors  in $U(1)$ anomaly cancellation  is here played
by the $tg$ and $cotg$  factors displayed in Table~\ref{betazn}.
Just like $\sin(2\pi kV_a)$ measures the mixing of the anomalous $U(1)$'s
with twisted moduli, in the present case those trigonometric factors
should measure the mixing of untwisted moduli $T_i$ with twisted moduli.
It would be interesting to confirm these couplings by a direct computation
in string theory.

\medskip

\subsection{Models with Wilson lines/non coincident branes}

Let us briefly comment on how the same anomaly cancellation mechanism works
in models with Wilson lines or with branes sitting at different points in
the compact space (both possibilities are related by T-duality). To keep
the discussion simple, we present two concrete examples with the $Z_3$
orientifold as starting point. In order to make the construction more
intuitive, we will perform a T-duality along the six compact dimensions,
thereby transforming the Wilson lines on the D9-branes into positions of
the T-dual D3 branes. The resulting models have orientifold group
$Z_3+Z_3 R_1 R_2 R_3 \Omega (-1)^{F_L}$ and contain 32 D3-branes and no
D7-branes.

\medskip

{\bf i) An example with a conformal subsector}

The first model we consider is analogous to that studied in \cite{finito}
\footnote{Even though the anomaly cancellation works analogously for the
model in \cite{finito}, the theory we consider is slightly more
illustrative for this particular issue.}. It is obtained upon placing 20
D3-branes at the origin, which is a $Z_3$ orientifold point, and
the remaining 12 D3-branes, in two groups of six, at two of the other
$Z_3$ fixed points (these are {\em orbifold} rather than {\em orientifold}
points), related by the orientifold projection. The spectrum is
\beqa
& SO(4) \times U(8) \times U(2)_1 \times U(2)_2 \times U(2)_3 \nonumber\\
& 3 [ (4,8;1,1,1) + (1,{\ov {28}};1,1,1) + (1,1;2,2,1) + (1,1;1,2,2) +
(1,1;2,1,2) ]
\label{wlfinito}
\eeqa

Notice that the model with all branes at the origin is continuously
connected to this theory, by moving 12 D3 branes off the origin to
the $Z_3$ orbifold points. Thus, it is expected the pattern of anomaly
cancellation will be similar in both models. In checking that this is so
we will learn an interesting bit or information concerning the behaviour
of twisted modes of orbifold (rather than orientifold) singularities in
the anomaly cancellation.

It is easy to check that the mixed $\sigma$-gauge anomalies with respect
to the different factors are
\beqa
{b^i}\,'_{SO(4)} = 6,\ {b^i}\,'_{SU(8)}=-3, \ {b^i}\,'_{SU(2)_a}=0, \
i=1,2,3
\eeqa

We see that the structure of $\sigma$-model anomalies reveals the
existence of two well-defined sectors. The modular anomalies related to
gauge groups for D3-branes at the origin ($SO(4)$ and $SU(8)$) are exactly
the same as in the model with all branes at the origin. On the other hand,
the triangle mixed anomalies with respect to the $SU(2)$ groups on the
D3-branes at the orbifold singularity vanish automatically, as expected
since this sector is conformal (see section 3.2).

It is straightforward to check that the anomalies for D3-branes at the
origin cancel through the GS mechanism exactly as in the model with
coincident branes. For the D3 branes at the orbifold singularity triangle
anomalies vanish, and no GS mechanism is required. We thus learn that
closed string twisted modes of orbifold (rather than orientifold)
singularities do not generate GS counterterms.

\medskip

{\bf ii) A further example}

The second example we would like to consider has appeared in \cite{lpt}.
In this model 23 D3-branes sit at the origin, and the remaining
9 D3-branes are stuck at different points fixed under $R_1 R_2 R_3 \Omega
(-1)^{F_L}$, i.e. O3-planes. Notice that this theory is {\em not}
continuously connected to the model with all branes at the origin, and so
anomaly cancellation is not obvious {\em a priori}.

The spectrum of the model is
\beqa
& SO(5)\times U(9) & \nonumber \\
& 3 \ [ \ (5,9) + (1,{\ov {36}}) \ ]&
\eeqa

Explicit computation of the triangle anomalies reveals that
\beq
{b^i}\,'_{SO(5)} = 6, \ {b^i}\,'_{SU(9)} = -3
\eeq
The anomaly is exactly cancelled by the GS mechanism, as discussed above.

Notice the D3-branes stuck at the O3-planes do not contribute any gauge
factors or matter multiplets, but this fact is not essential for the GS
mechanism to work. Actually, it is easy to construct related
models with several D3-branes at each O3-plane (at the expense of reducing
the number of D3-branes at the origin), so that the model contains
additional subsectors. The triangle anomalies with respect to these
gauge factors vanishes automatically, so no GS contribution is required.
This is expected since the O3-plane is {\em not} a fixed point with
respect to the $Z_3$ orbifold group and thus does not contain the
appropriate twisted fields. The triangle anomaly associated to D3-branes
at the origin is exactly as above and cancels through the usual GS
mechanism.

\medskip

We would like to stress that, even though we have discussed only two
simple examples, the mechanism for the cancellation of anomalies remains
valid for more complicated models.

\subsection{Mixed $\sigma $-model-gravitational anomalies}

The $\sigma$-model symmetries we have studied in the preceding sections
have mixed anomalies not only with the non-abelian gauge factors, but also
with gravity. In this section we address the cancellation of these
gravitational anomalies. For simplicity, we will restrict ourselves
to the case of odd order compact orientifolds, but the analysis holds in
general with suitable modifications of the relevant couplings.

The main difference between these anomalies and those studied above is
that the triangle anomalies also contain the contribution of massless
closed string states, running in the loop. Thus we will split the
triangle anomaly in two pieces:
\beq
b_{grav}^i\ =\ b_{closed}^i\ +\ b_{open}^i
\label{dos}
\eeq
The first contribution $b_{closed}^i$ has the form:
\beq
b_{closed}^i \ =\ 21\ +\ 1\ +\ \delta_T^i \ + \ \delta_M^i
\label{cerrado}
\eeq
As we discussed in the heterotic case, the $21+1$ come from the gravitino
and dilatino (partner of $S$) fields. The third term $\delta_T^i $
represent the contribution of the untwisted moduli fields themselves.
It is easy to check that $\delta_T^i= -3,-1$ for $Z_3$ and $Z_7$
respectively. The fourth term $\delta_M^i$ represent the contribution of
the twisted closed string states. Those have only non-linear transformations
with respect to the $SL(2,{\bf R})$ transformations, and so have
zero ``modular weights''. Their contribution is equal to the number of
twisted chiral fields.

This $b_{closed}^i$ piece of the anomaly is analogous to the one appearing
in the heterotic models, which also involves closed strings. As we
mentioned in section~2, in the heterotic case the contribution from the
field theory triangle diagrams is cancelled exactly by a one-loop diagram
which mixes the $S$ and $T_i$ fields.

The fact that the closed string contribution to the anomaly in the type I
case has the same structure suggest that this piece is cancelled in a
similar fashion. Namely, we propose that $b_{closed}^i$ is cancelled by a
one-loop mixing between the dilaton multiplet and the untwisted moduli
$T_i$. Thus, unlike what happens with mixed $U(1)$ or $\sigma $-model
anomalies for which the dilaton $S$ plays no role in anomaly
cancellation, in the case of mixed $\sigma $-model-gravity anomalies the
$S$ field gets transformed at {\it one loop } under an $SL(2,{\bf R})$
transformation.

The diagrammatic explanation for this behaviour is depicted in
Figure~\ref{diagclose}. The first two diagrams show the field theory
triangle anomaly due to type I closed string modes. The diagrams c) and d)
show the additional low-energy contributions corresponding to these
topologies. They are interpreted as the exchange of the dilaton multiplet
along the infinite tube, and its coupling to the untwisted moduli through
a one-loop subdiagram.

\begin{figure}
\centering
\epsfxsize=6.5in
\hspace*{0in}\vspace*{.2in}
\epsffile{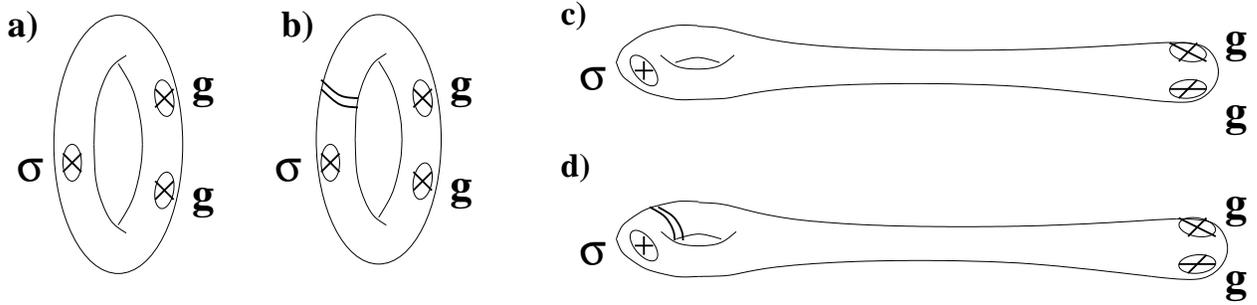}
\caption{\small The field-theory limits of the torus (a and c) and
Klein bottle (b and d) string world-sheets contributing to the mixed
$\sigma$-gravitational anomalies. The `cut' in the handles of diagrams b)
and d) represent a gluing that reverses the orientation. Diagrams a) and
b) represent closed string massless fields running though a loop, and
reproduce the field theory triangle anomaly. Diagrams c) and d) represent
the exchange of the dilaton multiplet along the infinite tube, and its
one-loop coupling to the moduli $T_i$. They provide the GS amplitudes
that cancel the triangle anomaly.}
\label{diagclose}
\end{figure}

Notice that this non-trivial transformation of the dilaton leads to no
contradiction with our previous results concerning $\sigma $-model-gauge
anomalies. This follows from the different dilaton dependence of the
$F\wedge F$ and $R\wedge R$ terms in Type I string theory. The
contribution arising from the coupling $SF\wedge F$ coupling upon the
one-loop transformation of the $S$-field is  a term of  higher order in
perturbation theory, as compared with the analogous coupling with gravity.
\footnote{This can be rephrased in string diagrammatics as follows.
A string topology combining the one-loop mixing of the $S$ and $T_i$ fields
and the coupling of $S$ to the gauge bosons will include one handle and one
boundary. If $g,b$ and $c$ are the number of handles, boundaries and
crosscaps, we will have a dilaton dependence with a power $(2-2g-b-c)=-1$,
whereas one-loop effects (like the diagrams in Fig.~\ref{diagmod}) have
a vanishing power ($g=0$, $b=2$, $c=0$, or $g=0$, $b=1$, $c=1$).}.
In summary, the triangle anomaly from closed string states
will be cancelled by the one-loop mixing of $S$ and $T_i$ fields.

\medskip

The remaining contribution, $b_{open}^i$, is on the other hand
cancelled through exchange of twisted closed string modes in the already
familiar fashion. The relevant diagrams providing the field theory
triangle anomaly and the GS terms are analogous to those involved in the
discussion of mixed $\sigma$-gauge anomalies (see figures~\ref{diagmod},
\ref{closmod}), with the difference that the graviton vertex operator
should be attached in the interior of the world-sheet. To support the
existence of this cancellation mechanism one can use techniques similar to
those used for mixed gauge anomalies. For the odd $Z_N$ we are considering,
using eqs.(\ref{specodd}) and (\ref{modgrav}) one gets
\beqa
b_{open}^i\ & =&\ - \sum_{a=1}^{(N-1)/2}w_a^2\ -\ {{w_0(w_0-1)}\over 2}\\
& - & {1\over 4}  \sum_{a=0}^{N-1} \ (w_a (w_{a+l_i}+w_{a-l_i})
+w_a(\delta_{2a+l_i}+\delta_{2a-l_i})\\
&+& \ {1\over 4} \sum_{j\not=i} w_a (w_{a+l_j}+w_{a-l_j})- \sum_{j\not=i}
w_a(\delta_{2a+l_j}+\delta_{2a-l_j})\ )
\label{bopen}
\eeqa
As before, doing an inverse discrete Fourier transform and
substituting the $w$'s one gets
\beq
b_{open}^i\  =\ {1\over {2N}}
\{ \ \sum _k   [\;\Tr\gamma_k \Tr\gamma_{-k}- Tr\gamma_{2k})
(\sum_{j\not=i}\cos(4\pi kv_j)  - \cos(4\pi kv_i)-1 \; ] \ \}
\label{bopeno}
\eeq
After some trigonometry one finds
\beq
b_{open}^i\ =\ {1\over{4 N}} \sum_{k=1}^{N-1} {\tilde {\omega}}_k^i
(\Tr\gamma_k \ \Tr\gamma_{-k}\ -\ \Tr\gamma_{2k})
\label{bopenplus}
\eeq
with
\beq
{\tilde {\omega}}_k^i\ =\ C_k\ \cotg(\pi kv_i)
\label{omega}
\eeq
Indeed, the result in eq.(\ref{bopenplus}) suggests the two contributions
from annulus and from Moebius strip. Also it shows an structure which is
compatible with its cancellation by the exchange of twisted closed string
massless states coupling simultaneously to untwisted moduli $T_i$ and
gravitons.

\section{Fayet-Iliopoulos terms and untwisted/twisted moduli mixing}

The presence and cancellation of $\sigma $-model anomalies have certain
implications for the structure of the effective low-energy action of Type
IIB compact orientifolds. We already observed how in the case of heterotic
orbifolds some mixing (eq.(\ref{ksgssigma})) must appear between the field
which transforms non-linearly under the $SL(2,R)_{T_i}$ symmetries (i.e.,
$S$ in the heterotic case) and the untwisted moduli $T_i$. Something
analogous must occur in the Type IIB orientifold case, but now it will
be the twisted fields $M_f^k$ which mix with the untwisted moduli. If, in
addition there are also anomalous $U(1)$ symmetries, as is generically the
case, the K\"ahler potential of the twisted fields will be of the form
\beq
K(M_f^k,{M_f^k}^*)\ =\ K(M_f^k+{M_f^k}^*
\ - \sum_a \delta_{GS \, k}^a  V_a  \ +
\  \sum_{i=1}^3 \delta_{GS\, k}^i \log(T_i+T_i^*) )
\label{kaememas}
\eeq
in order to have both $U(1)_a$ gauge invariance and $\sigma$-model
invariance with respect to the $i^{th}$ complex direction.  Thus the above
K\"ahler potential would be invariant respectively under the transformations
(we consider here the case of odd $N$ orientifolds for simplicity):
\beqa
Im M_f^k  & \rightarrow  &  Im M_f^k\ + \ \delta_{GS \, k}^f(a) \
\Lambda_a(x)\\ \nonumber
Im M_f^k &  \rightarrow &  Im M_f^k\ + \ \delta_{GS\, k}^i  \
\log(ic_iT_i+d_i)
\label{transfon}
\eeqa
where
\beq
\delta_{GS \, k}^f(a)\ =\ { w_a\, 2\, \sin(2\pi kV_a) } \ ;\
\delta_{GS\, k}^i\ =\  {2\tg(\pi k v_i) }
\label{lasdos}
\eeq
For a quadratic K\"ahler potential for the
$M_f^k$ fields, eq.(\ref{kaememas}) gives rise to a FI-term
corresponding to the $U(1)_a$ field :
\beq
\xi_a \ =\ -  \sum_f \sum_k \delta_{GS \, k}^a \ [\, M_f^k+{M_f^k}^*
\ +  \ \sum_{i=1}^3 \delta_{GS\, k}^i \log(T_i+T_i^*)\, ]
\label{fiagranel}
\eeq
This is an interesting result since it shows that in {\it compact} Type
IIB orientifolds the Fayet-Iliopoulos terms are controlled not only by the
blowing up modes of the orbifold singularities but also by the untwisted
moduli $T_i$. The FI-term in fact vanishes  at the points with:
\beq
2ReM_f^k
\ =  \ -  \sum_{i=1}^3 \delta_{GS\, k}^i \log(T_i+T_i^*) )
\label{punto}
\eeq
This corresponds to the SUSY-preserving vacuum when non-Abelian gauge
symmetry remains unbroken (FI-terms =0).

As we remarked in the previous section, the cancellation of mixed
$\sigma $-gravitational anomalies requires also the presence of a
mixing between the complex dilaton $S$ and the untwisted moduli,
very much analogous to the heterotic case. Thus one expects a form for
this mixing (again in the case of odd orientifolds):
\beq
K(S,S^*)\ =\ -\log(S+S^*  + \sum_i  \delta_{closed}^i \log(T_i+T_i^*)
)
\label{ksgssigma2}
\eeq

The additional untwisted moduli-dependence of FI-terms in compact
Type IIB , $D=4$, $N=1$ orientifolds have interesting implications
which will be discussed in more detail elsewhere \cite{ibappear}. Notice, 
for example, that it changes previous discussions  about matching of $Z_3$ 
and $Z_7$ orientifolds with their corresponding heterotic duals 
\cite{nilles} (for  example, for generic compact radii both the 
orientifold model and the  heterotic dual will have non-vanishing FI-terms 
associated to their  anomalous $U(1)$'s), or about the process of blowing 
up the orientifold singularities \cite{cvetic}.

\section{Conclusions}

In this paper we have addressed the issue of $\sigma$-model anomalies in
$D=4$, $N=1$ type IIB orientifold vacua. We have presented evidence
suggesting that anomalies associated to certain modular transformations
(those corresponding to complex planes rotated by all elements in the
orbifold group) are cancelled by a GS mechanism mediated by the exchange
of RR twisted closed string modes (for mixed gravitational anomalies,
the dilaton also plays a non-trivial role).

The main {\em a priori} reason to expect such cancellation is the duality
of these orientifolds with certain heterotic vacua, in which these
anomalies cancel. In heterotic models, this cancellation is required since
a discrete version of the $\sigma$-model symmetries corresponds to
T-duality, which is an exact symmetry of the full string theory.
It would be desirable to gain a better insight of $\sigma$-model
symmetries in type I string theory, in order to understand whether the
cancellation we have discussed follows as a consequence from a similarly
deep property of string theory.

The mechanism for the cancellation of $\sigma$-model anomalies that
we have uncovered is analogous to the GS mechanism which cancels $U(1)$
gauge anomalies, in that the exchanged fields are closed string twisted
modes. In particular, this has the consequence that the mixed anomalies
are allowed to be highly non-universal with respect to the different gauge
factors and gravity. This differs markedly from the behaviour in heterotic
vacua, and may be used to relax certain constraints on the low-energy
spectrum of phenomenologically interesting string vacua.

Finally we would like to stress that our conclusions have been based on a
detailed analysis of the triangle anomalies. In particular we have found
that rewriting the anomalies in terms of Chan-Paton traces is an extremely
useful trick which automatically exhibits the factorization properties of
the anomaly. In particular, it shows clearly the contribution of the RR
twisted modes to the GS mechanism to all these amplitudes, and, in the
case of the mixed gravitational anomaly, shows the necessity of having a
non-trivial transformation of the dilaton multiplet.

The anomaly cancellation mechanism requires the existence of interesting
tree-level mixings between the $T_i$ and the twisted closed string modes,
and a one-loop mixing between the $T_i$ and the dilaton. The consequences
of the existence of these couplings should be further explored. In
particular, they lead to an interesting modification for the FI terms for
the anomalous $U(1)$'s, which for compact models do not vanish at the
orbifold point. This fact had been overlooked in previous studies of the
consequences of the FI terms.

Type IIB orientifolds constitute an extremaly interesting set of models,
with properties often differing from the well-known behaviour of heterotic
vacua. As such, they are worthy of detailed exploration. We hope our
analysis helps in adding some useful information to our present knowledge
of the perturbative structure of type IIB orientifold vacua.

\centerline{\bf Acknowledgements}

We are thankful to G. Aldazabal, M. Klein, B.~Ovrut, E.~Poppitz and F. 
Quevedo  for useful discussions. A.~M.~U. is grateful to M.~Gonz\'alez
for encouragement and support, and to the Center for Theoretical Physics 
of M.~I.~T. for hospitality. L.E.I. and R.R. thank CICYT (Spain) and
the European Commission (grant ERBFMRX-CT96-0045) for financial support. 
The work of A.M.U. is supported by the Ram\'on Areces Foundation (Spain).

\newpage

\section{Appendix}

In this appendix we discuss the contribution of the different fixed points
to the cancellation of $U(1)$ anomalies in models with D5- and D9-branes.

For a given twist $k$ there are $C_k^2$ ($=\prod_{i=1}^3 4 \sin^2 \pi kv_i$)
fixed points. Let us label by an index $p$ the $4\sin^2 \pi k v_3$ fixed
points, located at the origin in the first two complex planes and  anywhere 
in the third. Since in our models the D5-branes sit at the origin in the 
two complex planes, it will be a combination of these twisted fields the 
one responsible for the cancellation of $U(1)$ anomalies in the D5-branes.
On the other hand, D9-branes fill the compact space completely and couple
to twisted modes from all the $C_k^2$ fixed points, which we label by an
index $f$. Let us define
\beqa
B_5^k & = & \frac{1}{2 \sin \pi k v_3} \sum_{p} M^k_{p} \nonumber \\
B_9^k & = & \frac{1}{\prod_{i=1}^3 2 \sin \pi k v_i} \sum_{f} M^k_{f}
\label{camposb}
\eeqa

Under a gauge transformation of the $a^{th}$ ($b^{th}$) $U(1)$ factor in
the sector of the D5-branes (D9-branes), with parameters $\Lambda^a_5$
($\Lambda^b_9$), the axion fields in $M^k_f$ transform as
follows
\beqa
Im\,M^k_{f} & \rightarrow & Im\, M^k_f + w_b\, 2\, \sin 2\pi k V^9_b \,
\Lambda^b_9 +  4\sin \pi k v_1 \sin \pi k v_2 \, w_a\, 2\, \sin 2\pi k V^5_a\,  \Lambda^a_5
\label{shiftinm}
\eeqa
(no sum in $a$, $b$ implied). Here the second contribution is only
present if the fixed point labeled by $f$ couples to the D5-branes.
This behaviour induces the following non-trivial transformation on the
fields (\ref{camposb})
\beqa
Im\, B_5^k & \rightarrow & Im\, B_5^k + C_{k}^{55} \, w_a\, 2\, \sin 2\pi k V^5_a\,
\Lambda^a_5  +
C_k^{59} \, w_b\, 2\, \sin 2\pi k V^9_b \, \Lambda^b_9 \nonumber \\
Im\, B_9^k & \rightarrow & Im\, B_9^k + C_{k}^{95} \, w_a\, 2\, \sin 2\pi k V^5_a\,
\Lambda^a_5 +
C_k^{99} \,w_b\, 2\, \sin 2\pi k V^9_b \,\Lambda^b_9
\label{shiftinb}
\eeqa
Notice the manifest symmetry between the couplings to D5- and D9-branes.
This exhibits the T-duality of this type of models.

The gauge kinetic functions for gauge fields on the D5- and D9-branes are
given by
\beqa
f^5_{a'} & = & T_3 + \frac{1}{N}\sum_{k=1}^{N/2} \cos 2\pi k V^5_{a'}
\; B^k_5 \nonumber \\
f^9_{b'} & = & S + \frac{1}{N}\sum_{k=1}^{N/2} \cos 2\pi k V^9_{b'}
\; B^k_9
\label{funcfs}
\eeqa
It is easy to check that the transformations (\ref{shiftinb}) then
generate the GS counterterms that cancel the triangle anomalies.

\end{document}